# A Binary Representation of the Genetic Code

Louis R. Nemzer[i]*

[1]Department of Chemistry and Physics, Halmos College of Natural Sciences and Oceanography, Nova Southeastern University, Davie, Florida, United States of America

*Corresponding author
E-mail: lnemzer@nova.edu (LRN)

# A Binary Representation of the Genetic Code

Louis R. Nemzer

*"The virtue of binary is that it's the simplest possible way of representing numbers. Anything else is more complicated."* - George Whitesides

**Abstract**

This article introduces a novel binary representation of the canonical genetic code based on both the structural similarities of the nucleotides, as well as the physicochemical properties of the encoded amino acids. Each of the four mRNA bases is assigned a unique 2-bit identifier, so that the 64 triplet codons are each indexed by a 6-bit label. The ordering of the bits reflects the hierarchical organization manifested by the DNA replication/repair and tRNA translation systems. In this system, transition and transversion mutations are naturally expressed as binary operations, and the severities of the different point mutations can be analyzed. Using a principal component analysis, it is shown that the physicochemical properties of amino acids related to protein folding also correlate with certain bit positions of their respective labels. Thus, the likelihood for a point mutation to be conservative, and less likely to cause a change in protein functionality, can be estimated.

**Author Summary**

This work introduces a new method of representing the genetic code using a binary system that reflects the relationships between nucleotide structures, as well as the amino acids they code for. Significant correlations are revealed between particular bits and the properties of the encoded amino acids. This paper also explores the way mutations can be classified as Boolean operations, and ranks the severity of the amino acid substitutions they cause. Thus, the binary labels are not arbitrary, but rather, have definite physiological meanings. This paper demonstrates a fruitful analogy between information represented as binary bits and the canonical genetic code. This connects the fields of information theory with molecular biology, since the inherent redundancy of the genetic code, an important source of error correction in the protein translation mechanism, relates to the risk of genetic disorders.

**Introduction**

Because of its central role in biological information processing, the canonical genetic code - which maps DNA codons onto corresponding amino acids - has been closely scrutinized for underlying symmetries. In this article, a novel binary representation of the code is introduced that accounts for both the chemical structures of the nucleotides themselves, as well as the physicochemical properties of the amino acids encoded. An accurate evaluation of the adaptive advantage of the code, which is robust to many point mutations and mRNA/tRNA mispairings, must consider not only the relatedness of amino acids separated by a single letter mutation, but also the probability of such a change or mispairing occurring in the first place based on the similarities of the nucleotides.

The primary addition this research makes to the existing literature is that the current work provides quantitative support for its novel binary classification system. That is, the choice of binary labels, as well as the order of the bits, have meaningful relationships with both the chemical structures of the nucleotides themselves, as well as the amino acids corresponding to codons in which they appear, as demonstrated with physicochemical data. This stands in contrast with many previous studies, which fixated on using the degree of degeneracy in the third letter as the primary or sole metric, and more crucially, treated the nucleotides as interchangeable labels for group theory analysis. This had the effect of strongly deemphasizing or obliterating entirely the physical reality of these biomolecules and their physicochemical similarities.

One of the first dichotomous divisions of the genetic code was due to theoretical physicist Yuri Rumer [1], [2], who noticed complete third-letter degeneracy in exactly half of the codon quartets [those of the form NCN or SKN. Refer to figures 1 and 2 for nucleotide abbreviations]. More recent generalizations [3] [4] [5] explored additional ways to bisect the genetic table. However, these works remained focused on classification according to the metric of third-letter degeneracy, along with hidden symmetries revealed by transformation rules involving the nucleotides. These rules show patterns under the interchange of nucleotides [6], while the current research takes account of the actual chemical properties of the nucleotides, as well as the amino acids they encode.

Modern computing, [7] which is built on the foundation of a binary system, provides a fertile analogy for the information conveyed by the quaternary encoding [8] of DNA. That is, each of the four possible nucleotide bases of DNA represents a maximum of $\log_2(4) = 2$ bits of information. However, this comparison extends far beyond a superficial similarity; the canonical genetic code, represented by a correspondence table between codons and amino acids, has a manifestly hierarchical organization [9] [10]. For example, the code distinguishes most clearly between pyrimidine (Y, Uracil or Cytosine) and purine (R, Adenine or Guanine) bases [11]. The number of heterocyclic rings differ in Y and R bases, and mutations that preserve this classification, called transitions, are more likely, but less damaging, than transversions between classifications. The binary identifiers chosen here to represent the nucleotide bases are not arbitrary. They are selected to reflect the molecular similarities exhibited by the nucleotides themselves. And, as will be shown, these labels have significant correlations with the physicochemical properties of the amino acids to which they correspond.

The system presented here accords with the theory that the genetic code has been shaped by natural selection [12] [13] [14], and that its evolution [15] [16] alongside the DNA mutation repair system [17] [18] and tRNA translation mechanism [19] has produced a table with the adaptive benefit [20] [21] that single-nucleotide mutations [22] most likely to cause a loss of protein function are also the most likely to be avoided [23] or fixed. Frameshifted "hidden" stop codons [24] can also quickly terminate protein synthesis upon ribosomal slippage. The protein translation machinery provides further robustness to error, in that non-cognate amino acids most likely to be misloaded by a tRNA molecule tend to have codons with the largest probability of being mismatched by the anticodon in the first place. [25] [26]

Ancestral versions of the genetic code may have already exhibited clustering of related amino acids as a result of stereochemical or biosynthetic similarities [27]. The inherent redundancy [28] in the code provides a measure of fault-tolerance [29], but also reduces the information [9] conveyed by each base. A fundamental irreducible representation [30] can be constructed as direct sum of two special unitary groups, one corresponding the Y/R dichotomy, and the other corresponding to W/S. Throughout this work, the "/" notation represents a binary choice between the two elements. For example, "Y/R" means "a choice between pyrimidine and purine." This "crystal basis model" [31] was presented to explain the observed minimization of the number of unique tRNA molecules required to complete protein translation by allowing "wobble pairing" [32] of certain similar codons to the same tRNA molecule. Hypothesized ancestral codes in which "expanded" codons had more than three bases [33] [34] may also lend themselves to binary representations [35].

Here, the standard amino acid correspondence table entries are recast as 6-bit binary messages. Due to the clustering of amino acids with similar physicochemical properties – the most critical [36] for proper protein folding and function being size, hydropathy [37], and charge – individual bit positions are correlated with specific properties. The classification system introduced here is not arbitrary; it places the most "determinative" bits first, and prioritizes the same nucleotide molecular features that nature does. This should be contrasted with some methods that attempt to solve the reverse problem – encoding binary data using DNA [38] – that implement an arbitrary revolving code in order to minimize the occurrence of repeated bases, irrespective of the structures of the nucleotides. Following conventional codon tables, the system introduced here focuses on mRNA, so it uses uracil instead of thymine, but since these bases differ only by a single methyl group, it is likely that the same or similar physicochemical properties that are recognized by the mRNA-to-peptide translation machinery are also utilized by the DNA replication and repair mechanisms.

## Method

A set of four elements, such as the nucleotides, can be divided into $\binom{4}{2} = 6$ unique duos. Only two bits are needed to unambiguously identify each element, which leaves some freedom of choice in labeling systems. Here, each of the four nucleotide bases is assigned a 2-bit identifier (Figure 1) in which the most meaningful molecular similarities are emphasized. The first bit is a 0 for the pyrimidine bases (Y, two heterocyclic rings), and 1 for the purines (one heterocyclic ring). The second bit is a 0 for the "weak" bases that form 2 hydrogen bonds with each other during Watson-Crick pairing (W = U or A), and 1 for the "strong" bases that form 3 hydrogen bonds (S = C or G). Thus, the code for U is 00, C is 01, A is 10, and G is 11. The bases can also be paired a third way – into keto (K = U or G) or amino (M = A or C). [5] The keto bases have both bits as either 1 or 0, so the XOR operation applied its bits would give 0, while the amino bases have dissimilar bits (XOR = 1).

|          | $Y_{0-}$  | $R_{1-}$  |
|----------|-----------|-----------|
| $W_{-0}$ | $U_{00}$  | $A_{10}$  |
| $S_{-1}$ | $C_{01}$  | $G_{11}$  |

**Figure 1:** Nucleotide bases and their 2-bit identifiers as subscripts, along with the IUPAC letter abbreviations for duos. The four bases are assigned a binary identifier in which the first bit designates whether it is pYrimidine (0-) or puRine(1-). The second bit shows if the base is Weak (-0), forming two hydrogen bonds during Watson-Crick pairing, or Strong (-1), forming three. These pairings were chosen to prioritize the same physiological characteristics most recognized by the DNA repair and amino acid translation systems.

A summary of the labeling system is given by the truth table (Figure 2). Each of the six duos contains two of the nucleotides, and every nucleotide is a member of exactly three duos - one of each complimentary set Y/R, W/S, and K/M - indicating a similarity it shares with each of the three other nucleotides.

|     | U  | C  | A  | G  |
|-----|----|----|----|----|
|     | 00 | 01 | 10 | 11 |
| Y 0- | 1 | 1 | 0 | 0 |
| R 1- | 0 | 0 | 1 | 1 |
| W -0 | 1 | 0 | 1 | 0 |
| S -1 | 0 | 1 | 0 | 1 |
| K 0 XOR | 1 | 0 | 0 | 1 |
| M 1 XOR | 0 | 1 | 1 | 0 |

**Figure 2:** Nucleotide truth table. The four nucleotides (U, C, A, G) are listed according to their respective 2-bit identifiers. Each base can join with each of the three others to make a duo based on physicochemical similarities. Complimentary duos (Y vs. R, then W vs. S, then K vs. M) are ordered according to their physiological relevance.

To further illustrate the molecular basis of this classification system, Figure 3 shows the four bases during Watson-Crick pairing. Here, the hierarchical nature of this classification system is manifest. That is, the number of heterocyclic rings (Y vs. R) is the most salient feature. This corresponds to the well-established finding that transitions among Y or R bases are more common than transversions between them. The next most relevant feature is the number of hydrogen bonds, in that U and A pair with each other using 2 hydrogen bonds (Weak), while C and G pair with three (Strong). Finally, of the possible pairings, the least important, from a physiological viewpoint, is the presence of an amino or keto group attached at the C6 (for the purines) or C4 (for the pyrimidines) position. These features are represented schematically in Figure 4.

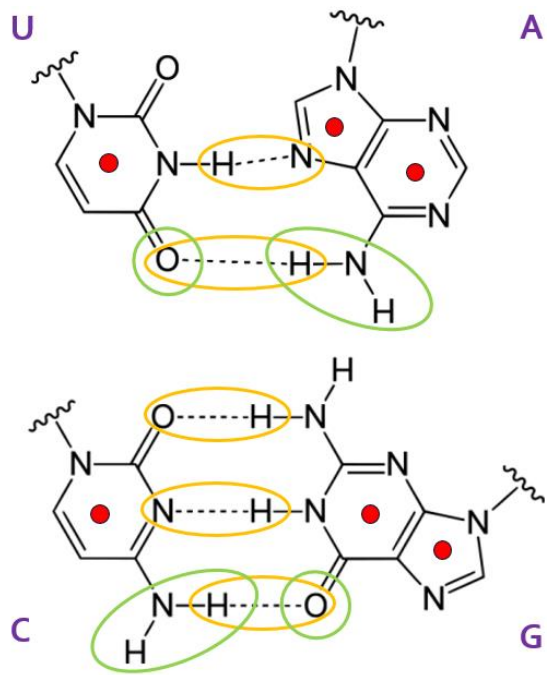

**Figure 3:** An illustration of the physical meaning of the binary identifiers based on the molecular structures of the nucleotide bases. Heterocyclic rings are marked with red circles, hydrogen bonds with yellow ellipses, and the amino or keto groups in question in green ellipses. Y (R) bases have 1 (2) heterocyclic rings. The weak (strong) bases pair with each other using 2 (3) hydrogen bonds. In the keto (amino) bases, the named group acts as one of the hydrogen bond donors (acceptors).

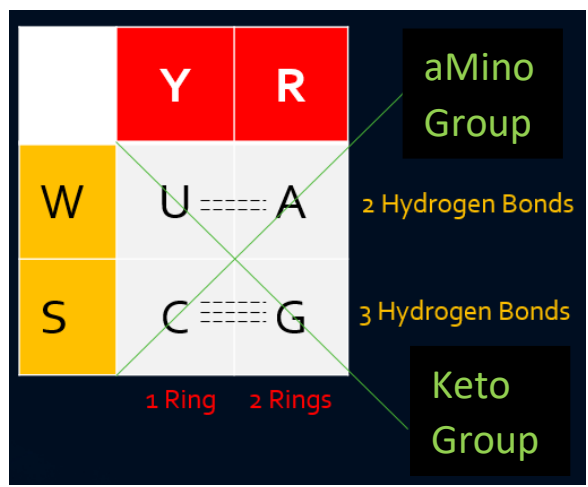

**Figure 4:** A schematic representation of the six nucleotide duos that captures the essential physiological similarities of each pairing. The dashed lines represent the hydrogen bonds formed between U and A and between C and G during Watson-Crick pairing.

The basic unit of mRNA-to-protein translation is the trinucleotide codon. The organization of conventional tables, which indicate the amino acid corresponding to each codon, reflects the long-established finding that the second letter of each codon conveys the most information about the intended amino acid, followed by the first letter [9]. So, related amino acids tend to be grouped into the same vertical column. The third letter of a codon is often degenerate, in that it does not change the identity of the encoded amino acid once the first two are known. Thus, to prioritize the most significant bits, the classification system introduced here reorders the nucleotides of the codon to be 2, 1, 3. Figure 5 provides an example of the binary representation using the codon **AUG**, which codes for the amino acid methionine. The 6-bit index for the codon is determined by concatenating the 2-bit identifiers from the second nucleotide ($U_{00}$), the first nucleotide ($A_{10}$), and then the third nucleotide ($G_{11}$), yielding **001011**. This method is equivalent to the following series of questions: Is the second base a purine? (0 for No, 1 for Yes). Is the second base strong? (0 for No, 1 for Yes). The questions are repeated for the first, and then third base of the codon.

| BIT | | | | | |
|---|---|---|---|---|---|
| 1 | 2 | 3 | 4 | 5 | 6 |
| 2nd Base R? | 2nd Base S? | 1st Base R? | 1st Base S? | 3rd Base R? | 3rd Base S? |
| NO **0** | NO **0** | YES **1** | NO **0** | YES **1** | YES **1** |
| $U_{00}$ A | | $A_{10}$ U | | $G_{11}$ G | |
| Met | | | | | |

**Figure 5**: Example showing the method for determining the 6-bit index of each codon. Here, the codon **AUG**, which corresponds to the amino acid methionine, has the index **001011**. The second letter of the codon is listed first, followed by the first and third letters.

Following this method, each of the 64 codons is assigned a unique 6-bit index that places the most important information first. A complete amino acid correspondence table under this method is provided as Figure 6. Another representation of the system, which casts the table as a binary (dichotomic) search tree, is given in Appendix Figure 1. Some previously identified clusterings of important amino acid properties on the table can now be recast as binary properties. For examples, all of the charged amino acids have 1 as the first bit, and all amino acids with indices that start with 00 are hydrophobic. These relationships, and others, are tested systematically in the sections that follow.

| 1st Base | 2nd Base | | | | | | | | | | | | 3rd Base |
|---|---|---|---|---|---|---|---|---|---|---|---|---|---|
| | U | | | C | | | A | | | G | | | |
| U | UUU | 000000 | Phe | UCU | 010000 | Ser | UAU | 100000 | Tyr | UGU | 110000 | Cys | U |
| | UUC | 000001 | Phe | UCC | 010001 | Ser | UAC | 100001 | Tyr | UGC | 110001 | Cys | C |
| | UUA | 000010 | Leu | UCA | 010010 | Ser | UAA | 100010 | Stp | UGA | 110010 | Stp | A |
| | UUG | 000011 | Leu | UCG | 010011 | Ser | UAG | 100011 | Stp | UGG | 110011 | Trp | G |
| C | CUU | 000100 | Leu | CCU | 010100 | Pro | CAU | 100100 | His | CGU | 110100 | Arg | U |
| | CUC | 000101 | Leu | CCC | 010101 | Pro | CAC | 100101 | His | CGC | 110101 | Arg | C |
| | CUA | 000110 | Leu | CCA | 010110 | Pro | CAA | 100110 | Gln | CGA | 110110 | Arg | A |
| | CUG | 000111 | Leu | CCG | 010111 | Pro | CAG | 100111 | Gln | CGG | 110111 | Arg | G |
| A | AUU | 001000 | Ile | ACU | 011000 | Thr | AAU | 101000 | Asn | AGU | 111000 | Ser | U |
| | AUC | 001001 | Ile | ACC | 011001 | Thr | AAC | 101001 | Asn | AGC | 111001 | Ser | C |
| | AUA | 001010 | Ile | ACA | 011010 | Thr | AAA | 101010 | Lys | AGA | 111010 | Arg | A |
| | AUG | 001011 | Met | ACG | 011011 | Thr | AAG | 101011 | Lys | AGG | 111011 | Arg | G |
| G | GUU | 001100 | Val | GCU | 011100 | Ala | GAU | 101100 | Asp | GGU | 111100 | Gly | U |
| | GUC | 001101 | Val | GCC | 011101 | Ala | GAC | 101101 | Asp | GGC | 111101 | Gly | C |
| | GUA | 001110 | Val | GCA | 011110 | Ala | GAA | 101110 | Glu | GGA | 111110 | Gly | A |
| | GUG | 001111 | Val | GCG | 011111 | Ala | GAG | 101111 | Glu | GGG | 111111 | Gly | G |

**Figure 6:** The standard amino acid correspondence table with 6-bit indices included. Refer to appendix table 2 for a graphical interpretation of the binary encoding method.

A graphical interpretation [39] of the binary encoding is given in appendix table 2. In addition to organizing the information conveyed by codons, another benefit of using a binary representation is that mutations can be considered as Boolean operations [40]. Starting with a particular nucleotide, a mutation to each of the three other bases can be characterized according to the classification it *preserves*, that is, the duo formed by the original and mutated base. Following this, a transition mutation between U and C would be classified as Y, while a transversion between U and A, or U and G, would be W or K, respectively (see Figure 7). Using the binary identifiers, Y and R mutations flip the value of the second bit, from 0 to 1, or vice versa, W and S mutations flip the value of the first bit, and K and M mutations flip the value of both bits. Note parenthetically that if the body did not recognize a hierarchy of molecular similarities and all mutations had an equal inherent likelihood, simple probability would dictate that transversion mutations – which can be W, S, K, or M – would be twice as likely as transitions, which can only be Y or R. In fact, transitions are observed to be about three times *more* common than transversions, implying a DNA replication and repair system that results in sixfold lower fidelity when distinguishing bases belonging to the same Y or R duo, compared with those that are members of different duos.

In addition, this new system can classify specific mutation mechanisms. For example, a "CpG" transition mutation [41] can occur when a C base, which is followed by a G, becomes methylated as an epigenetic mark. If that C subsequently loses its amino group and replaces it with a carbonyl – a Y mutation – it will become a thymine base, the DNA analogue of U. In this way, the base maintains its membership as a pyrimidine, but switches its other two classifications: from amino to keto, and from strong to weak.

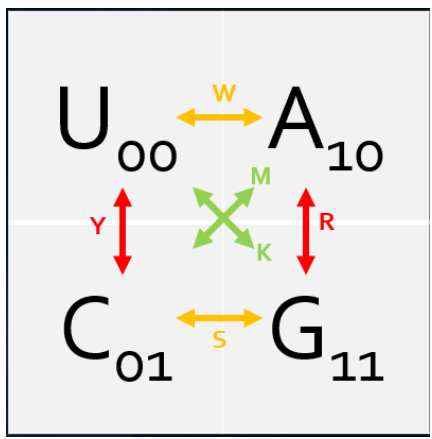

**Figure 7:** Mutation nomenclature. Each base has three mutation possibilities, denoted here by the duo it shares with the new base. Exactly one duo classification is preserved, while the other two are inverted. For example, a mutation from U to C (or vice versa), is a Y mutation, since they are both pyrimidines. This will have the effect of reversing the W or S, as well as the K or M, identities. A Y or R mutation preserves the value of the first bit, and flips the second from 0 to 1, or from 0 to 1. On the other hand, W or S mutations preserve the second bit and flip the first. K or M mutations flip both bits. Y or R mutations, which are pyrimidine to pyrimidine and purine to purine changes, are called transitions, and W,S, K, or M mutations (pyrimidine to purine, or vice versa) are transversions.

To demonstrate the value of the binary representation, all possible single nucleotide mutations were classified by type and graded according to the severity of the resulting change in the amino acid indicated by the codon. Mutations to or from stop codons were omitted. The BLOSUM62 substitution matrix [42] [43], which compares evolutionarily divergent proteins to see how often one amino acid replaces another, was used as the measure of mutation severity. This substitution matrix was chosen, since, as opposed to others like PAM, it is less endogenously biased [44] by single mutation likelihoods.

A principal component analysis (PCA) was used to systematically quantify the relationship between codon placement in the table, the 6-bit indices, and the physicochemical properties of the encoded amino acids. Briefly, PCA is a method for summarizing data when some of the characteristics are expected to be correlated. This can be thought of as taking the n-dimensional data matrix and performing a rotation of the axes so the first principal component (PC) is in the direction of the highest variance. The next component is in the direction orthogonal to the first component that captures the g remaining variance. This process is repeated until all n directions are assigned. The PC directions can be expressed as a linear combination of the original axes, however, all but the first few principal components are generally neglected, reducing the dimensionality of the data set but maintaining the majority of its information. In practice, the PCs are usually computed as the eigenvectors of the standardized data covariance matrix. To measure the physiological import of the bit positions, 81 separate ANOVA tests were run.

**Results**

Figure 8 shows the fraction of each type of mutation that causes an amino acid substitution of a particular severity, with smaller BLOSUM62 values corresponding to more damaging changes. A very low BLOSUM62 score means that the substitution is especially unsuitable for maintaining correct protein folding, such as replacing a hydrophobic amino acid with a hydrophilic one, or vice versa. Conversely, the largest values correspond to silent mutations that preserve highly conserved amino acids that have particular properties related to native protein conformation. For example, the ability for cysteine residues to form disulfide bridges make them very resistant to substitution. The rightmost bin of the chart, with BLOSUM62 scores from 6 to 9, shows that over 10% of transitions (Y or R) correspond to a silent mutation in a strongly conserved amino acid. A significantly smaller fraction of both kinds of transversion (W or S, and K or M) do this. The remaining transitions are about evenly distributed among the other three bins. In contrast, the fraction of both W or S and K or M transversions only increases as the severity worsens. Nearly half of each kind, 45% of W or S mutations and 49% of K or M mutations, represent strongly disfavored amino acid substitutions with negative BLOSUM62 scores.

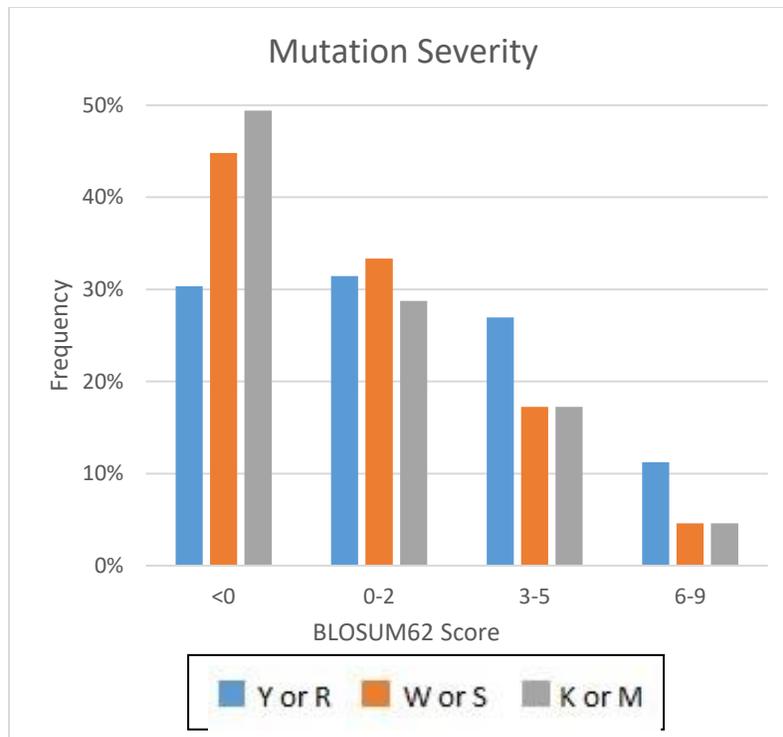

**Figure 8:** Mutation severity frequency by type of mutation, using the BLOSUM62 substitution matrix. Transition mutations are between pyrimidine (Y) or between purine (R) bases, while a transversion preserves exactly of one of these classifications: weak (W), strong (S), amino (M), or keto (K). Larger BLOSUM62 values correspond to amino acid substitutions more likely to be found when comparing evolutionary divergent proteins. The rightmost bin corresponds to silent mutations that leave critical amino acids unchanged. On the other hand, the leftmost bin contains substitutions that are particularly damaging to the proper folding of the protein.

For the principal component analysis, physicochemical data (Figure 9) for six amino properties were used: molecular weight, van der Waals volume (the space the amino acid occupies), solvent-accessible surface area, hydropathy, isoelectric point (which determines charge at physiological pH), and the tendency to be buried in the interior of proteins, as opposed to the surface, called "buriability". It is well known that the most critical determinate of normal protein function, the ability to fold into the native conformation, is driven by hydrophobic, steric, and to a lesser extent, Columbic forces. These six properties, which have some clear correlations with each other, were chosen in light of their critical importance for proper amino acid folding and function. The data correlation matrix is shown as Figure 10. As expected, molecular weight, van der Waals volume, and surface area form a correlated group based on amino acid size - while buriability is driven primarily by hydropathy. The isoelectric point is more ambiguously related to hydropathy, since both positively and negatively charged amino acids will both be strongly hydrophilic, while uncharged residues may be polar or nonpolar.

| Abv | Letter | Mol Weight | VdW Vol | Area | Hydrophobic | pI | Buriability |
|---|---|---|---|---|---|---|---|
| ALA | A | 89.09 | 67 | 115 | 1.8 | 6 | 38 |
| CYS | C | 121.16 | 86 | 135 | 2.5 | 5.07 | 47 |
| ASP | D | 133.1 | 91 | 150 | -3.5 | 2.77 | 14.5 |
| GLU | E | 147.13 | 109 | 190 | -3.5 | 3.22 | 20 |
| PHE | F | 165.19 | 135 | 210 | 2.8 | 5.48 | 48 |
| GLY | G | 75.07 | 48 | 75 | -0.4 | 5.97 | 37 |
| HIS | H | 155.15 | 118 | 195 | -3.2 | 7.59 | 19 |
| ILE | I | 131.17 | 124 | 175 | 4.5 | 6.02 | 65 |
| LYS | K | 146.19 | 135 | 200 | -3.9 | 9.74 | 4.2 |
| LEU | L | 131.17 | 124 | 170 | 3.8 | 5.98 | 41 |
| MET | M | 149.21 | 124 | 185 | 1.9 | 5.74 | 50 |
| ASN | N | 132.12 | 96 | 160 | -3.5 | 5.41 | 10 |
| PRO | P | 115.13 | 90 | 145 | -1.6 | 6.3 | 24 |
| GLN | Q | 146.14 | 114 | 180 | -3.5 | 5.65 | 6.3 |
| ARG | R | 174.2 | 148 | 225 | -4.5 | 10.76 | 0 |
| SER | S | 105.09 | 73 | 115 | -0.8 | 5.68 | 24 |
| THR | T | 119.12 | 93 | 140 | -0.7 | 5.6 | 25 |
| VAL | V | 117.15 | 105 | 155 | 4.2 | 5.96 | 56 |
| TRP | W | 204.23 | 163 | 255 | -0.9 | 5.89 | 23 |
| TYR | Y | 181.19 | 141 | 230 | -1.3 | 5.66 | 13 |

**Figure 9:** Physicochemical data used in the principal component analysis. The categories are molecular weight [45], van der Waals volume [46], solvent-accessible surface area [47], hydrophobicity index [48], isoelectric point [45], and tendency for the amino acid to be buried in the interior of proteins [49].

|  | Weight | Vol | Area | Hydro | pI |
|---|---|---|---|---|---|
| Vol | 0.934 | | | | |
| Area | 0.978 | 0.972 | | | |
| Hydro | -0.271 | -0.080 | -0.218 | | |
| pI | 0.205 | 0.371 | 0.292 | -0.203 | |
| Bury | -0.336 | -0.178 | -0.293 | 0.945 | -0.290 |

**Figure 10:** Data correlation matrix. Molecular weight, van der Waals volume, and surface-accessible area are closely related. Buriability is strongly related to hydrophobicity, and, to a lesser extent, small size. The isoelectric point is not monotonically related to hydropathy or buriability, since both positively (basic) and negatively (acidic) charged amino acids tend to be hydrophilic. However, basic residues with high pI values do tend to be large in order to stabilize their excess positive charge.

The results of the PCA are displayed in Figure 11. The power of each component represents how much of the overall variance in the data it captures. The first three components collectively account for 98.6%, so the last three can be safely discarded. The projection of each component onto the original data axes is also shown.

| Power | Name | Mol Weight | Volume | Area | Hydropathy | pI | Buriability |
|---|---|---|---|---|---|---|---|
| **55.6%** | **PC1** | **-0.509*** | **-0.490*** | **-0.513*** | 0.273 | -0.250 | 0.316 |
| **28.6%** | **PC2** | -0.207 | -0.328 | -0.245 | **-0.641*** | 0.098 | **-0.608*** |
| **14.4%** | **PC3** | -0.230 | 0.027 | -0.115 | 0.166 | **0.948*** | 0.089 |
| **0.8%** | **PC4** | 0.159 | -0.194 | 0.077 | **-0.642*** | 0.099 | **0.713*** |
| **0.4%** | **PC5** | **0.728*** | **-0.595*** | -0.154 | 0.249 | 0.141 | -0.104 |
| **0.2%** | **PC6** | -0.300 | **-0.509*** | **0.796*** | 0.113 | 0.024 | -0.059 |

**Figure 11:** Principal component analysis, in which coefficients with magnitudes larger than 0.4 are starred. The six physicochemical properties divide into three groups: *Size*, which includes weight, volume, and area; *hydropathy*, which is connected with buriability; and *isoelectric point*. Since its coefficients for molecular weight, volume, and area are negative, PC1 should interpreted as *small*, as larger values of PC1 correspond to smaller amino acids. Higher PC2 values mean more *hydrophilic* residues, which are expected to be harder to bury. Finally, PC3 corresponds to *positively* charged (basic) amino acids with high isoelectric points (pI). The remaining principal components can be neglected, since they collectively capture only 1.4% of the variation in the data.

Again, the six physicochemical properties divide naturally into three groups. The first principal component, PC1, is inversely related to molecular weight, surface-accessible volume, and van der Waals area. Thus, it can be identified with the general descriptor "small." Next, PC2 is highest for the "hydrophilic" residues, and inversely related to buriablity. Finally, a high isoelectric point is indicative of a large value for PC3, "positive," or basic amino acids. The PC values for all 20 amino acids is given in Figure 12.

| Letter | PC1 | PC2 | PC3 | PC4 | PC5 | PC6 | Middle | Type |
|---|---|---|---|---|---|---|---|---|
| A | 2.530 | 0.294 | 0.623 | 0.180 | 0.066 | 0.253 | C | Nonpolar |
| C | 1.794 | -0.745 | -0.067 | -0.038 | 0.296 | -0.043 | G | Polar |
| D | 0.556 | 1.261 | -1.911 | -0.002 | -0.078 | -0.127 | A | Charged |
| E | -0.418 | 0.584 | -1.837 | -0.262 | -0.252 | 0.138 | A | Charged |
| F | -0.654 | -2.088 | -0.306 | -0.067 | 0.118 | 0.051 | U | Nonpolar |
| G | 3.337 | 1.330 | 0.673 | -0.237 | 0.086 | -0.087 | G | Nonpolar |
| H | -1.367 | 0.612 | 0.451 | -0.393 | 0.113 | 0.067 | A | Charged |
| I | 0.874 | -2.433 | 0.495 | -0.222 | -0.246 | -0.046 | U | Nonpolar |
| K | -2.185 | 1.207 | 1.561 | 0.059 | -0.268 | -0.002 | A | Charged |
| L | 0.464 | -1.469 | 0.332 | 0.563 | -0.155 | -0.087 | U | Nonpolar |
| M | 0.004 | -1.575 | -0.032 | -0.299 | -0.011 | -0.093 | U | Nonpolar |
| N | -0.079 | 1.449 | -0.533 | 0.045 | -0.003 | 0.027 | A | Polar |
| P | 0.767 | 0.897 | 0.277 | -0.062 | -0.078 | 0.063 | C | Nonpolar |
| Q | -0.946 | 1.175 | -0.563 | 0.187 | -0.072 | -0.045 | A | Polar |
| R | -3.431 | 1.056 | 1.792 | -0.067 | 0.093 | -0.042 | G | Charged |
| S | 1.732 | 1.119 | 0.127 | 0.136 | 0.157 | -0.066 | C/G | Polar |
| T | 0.907 | 0.600 | -0.057 | 0.141 | -0.014 | -0.098 | C | Polar |
| V | 1.488 | -1.655 | 0.542 | 0.044 | -0.096 | 0.076 | U | Nonpolar |
| W | -3.123 | -1.281 | -0.795 | -0.016 | 0.171 | -0.054 | G | Nonpolar |
| Y | -2.253 | -0.337 | -0.771 | 0.312 | 0.172 | 0.114 | A | Polar |

**Figure 12:** Principal component analysis values. The middle nucleotide and amino acid type is also indicated. Acidic and basic amino acids are grouped together under the classification "charged."

It has previously been shown [50] that, when plotted in PCA space [51], amino acids cluster according to the middle nucleotide of their corresponding codons, and this feature is readily apparent here. In Figure 13a, all amino acids are plotted according to the first three principal component values. The same information is also represented with a 3D-printed physical model in Figure 13b. The projection on the PC1-PC2 plane is Figure 13c.

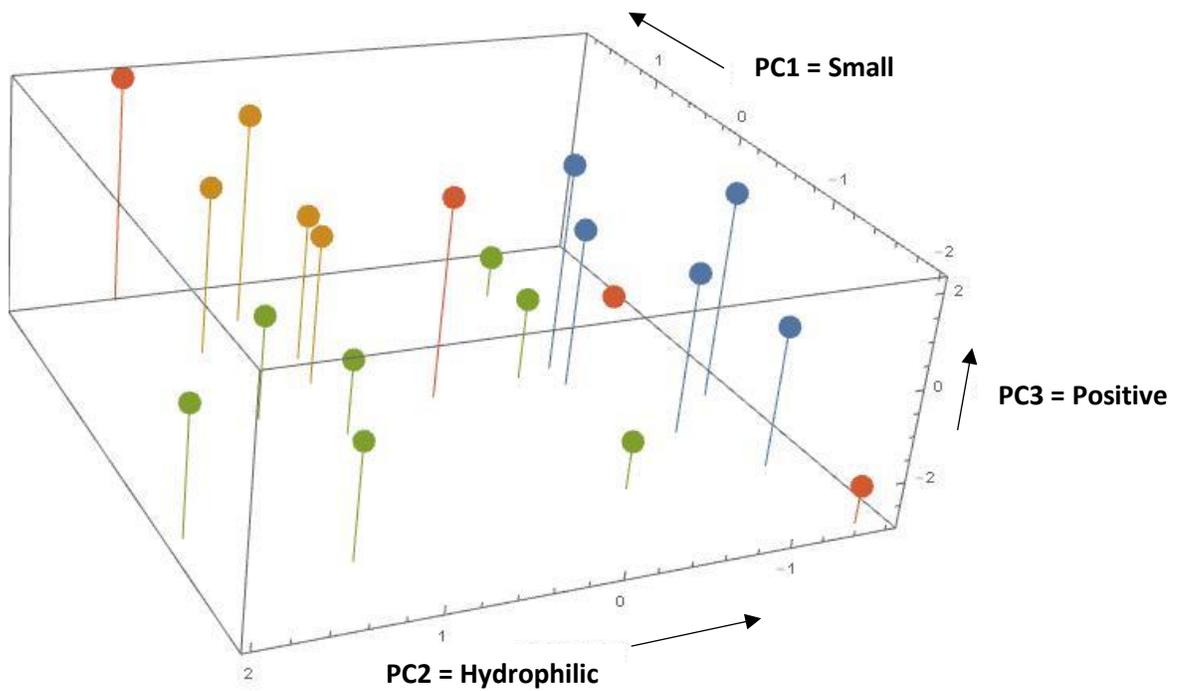

**Figure 13a:** Amino acids in physicochemical space. PC1 is identified as "small," PC2 is "hydrophilic," and PC3 is "positive." The colors indicate the middle letter of the codons that correspond to each amino acid. (Green = A, Blue = U, Yellow = C, Red = G)

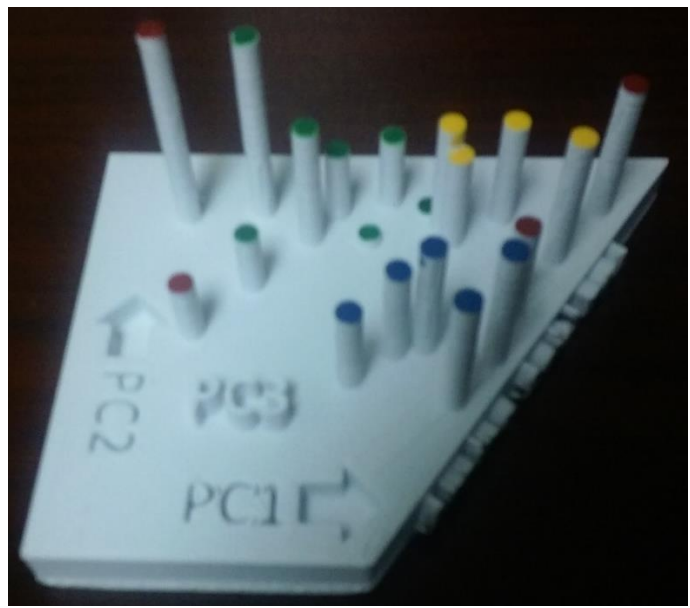

**Figure 13b:** 3D-Printed model of the data in the previous Figure.

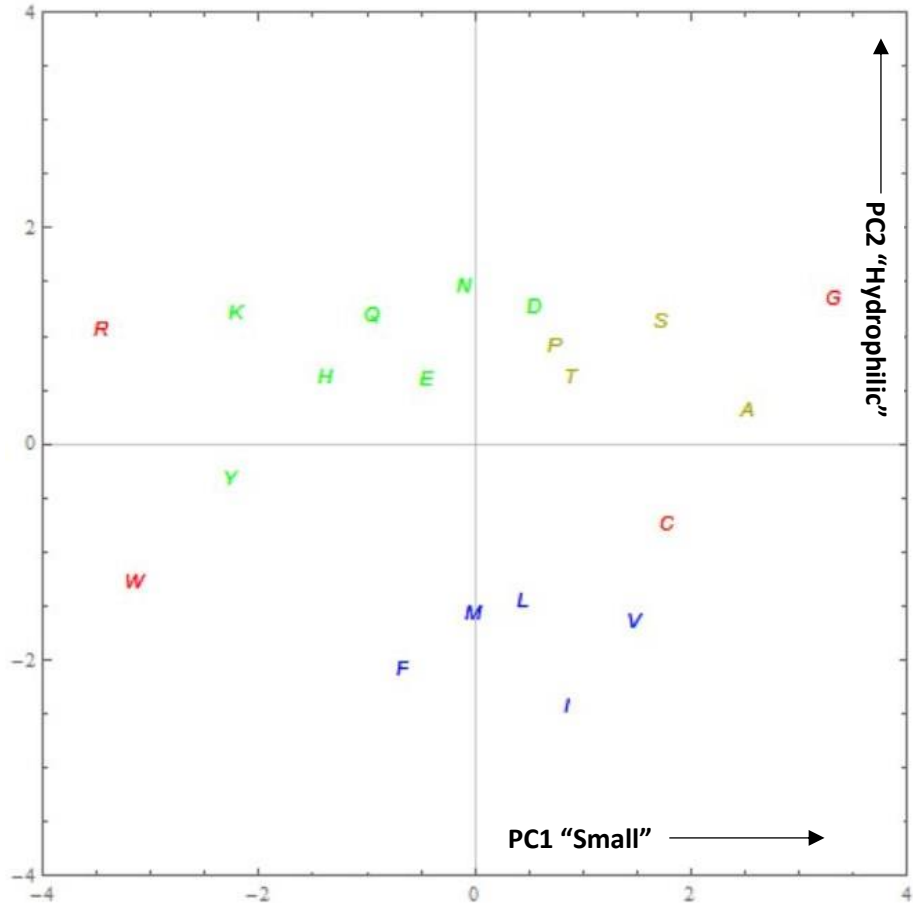

Figure **13c:** Projection of the PCA data on the PC1-PC2 plane.

The results of the ANOVA tests that measure the variance of the raw physicochemical data and first three principal components attributable to each bit position is given in Figures 14 and 15, respectively. For each test, both the F-Statistic, which is the ratio of the intergroup variability to the intragroup variability, and p-value are given, although this analysis will focus on the former. Each bit, or XOR operation of two bits, indicates a duo to which a particular nucleotide of the codon belongs. A high correlation with a bit position reflects the sensitivity of a physicochemical property or principal component to the corresponding duos. For example, the largest F-Statistic occurs for the connection between hydropathy and bit 1, which gives the Y/R identity of the second nucleotide of the codon (denoted YR2). This reflects the trend that many hydrophobic amino acids occupy the left side (bit 1 = 0) of the table, while charged residues are confined to the right. In fact, the first two bits of an index, which together determine the middle nucleotide, are by far the most informative, relating significantly to all six raw physicochemical categories. Among the last four bits of the 6-bit index, which stand for the first and third letters, only XOR(3,4), which gives the K/M identity of the first nucleotide of the codon, shows significant associations. This reflects the large degeneracy of the third base, especially regarding WS3 and KM3.

| BIT | DUO | Mol Weight | | Volume | | Area | | Hydropathy | | pI | | Buriability | |
|---|---|---|---|---|---|---|---|---|---|---|---|---|---|
| | | F | P | F | p | F | p | F | p | F | p | F | p |
| 1 | YR2 | 0.013 | 0.910 | 0.459 | 0.501 | **25.011** | 0.000 | **45.015** | 0.000 | 1.192 | 0.279 | **35.914** | 0.000 |
| 2 | WS2 | **42.057** | 0.000 | 5.188 | 0.026 | 0.187 | 0.667 | **15.836** | 0.000 | **11.013** | 0.002 | 0.519 | 0.474 |
| XOR(1,2) | KM2 | 0.002 | 0.968 | **28.803** | 0.000 | 1.009 | 0.319 | **12.825** | 0.001 | 4.119 | 0.047 | 2.885 | 0.095 |
| 3 | YR1 | 2.027 | 0.160 | 4.289 | 0.043 | 6.738 | 0.012 | 0.040 | 0.843 | 6.518 | 0.013 | 1.110 | 0.296 |
| 4 | WS1 | 2.334 | 0.132 | 1.547 | 0.218 | 0.551 | 0.461 | 0.732 | 0.396 | 1.052 | 0.309 | 0.020 | 0.887 |
| XOR(3,4) | KM1 | 0.942 | 0.336 | 5.774 | 0.019 | **12.680** | 0.001 | 3.166 | 0.080 | **8.087** | 0.006 | **18.644** | 0.000 |
| 5 | YR3 | 0.328 | 0.569 | 0.748 | 0.390 | 1.030 | 0.314 | 0.062 | 0.805 | 1.166 | 0.285 | 1.792 | 0.186 |
| 6 | WS3 | 0.134 | 0.715 | 0.015 | 0.902 | 0.074 | 0.787 | 0.005 | 0.945 | 0.050 | 0.824 | 0.016 | 0.900 |
| XOR(5,6) | KM3 | 0.134 | 0.715 | 0.015 | 0.902 | 0.074 | 0.787 | 0.005 | 0.945 | 0.050 | 0.824 | 0.016 | 0.900 |

**Figure 14a:** Results of ANOVA analysis for each bit position with the six physicochemical categories. The affected nucleotide position and duo is also given. For example, bit 1 determines whether the second letter of the codon is Y or R. Each entry lists the F-Statistic along with the p-value, and those with an F-statistic exceeding 7 (and have p < 0.01) are highlighted.

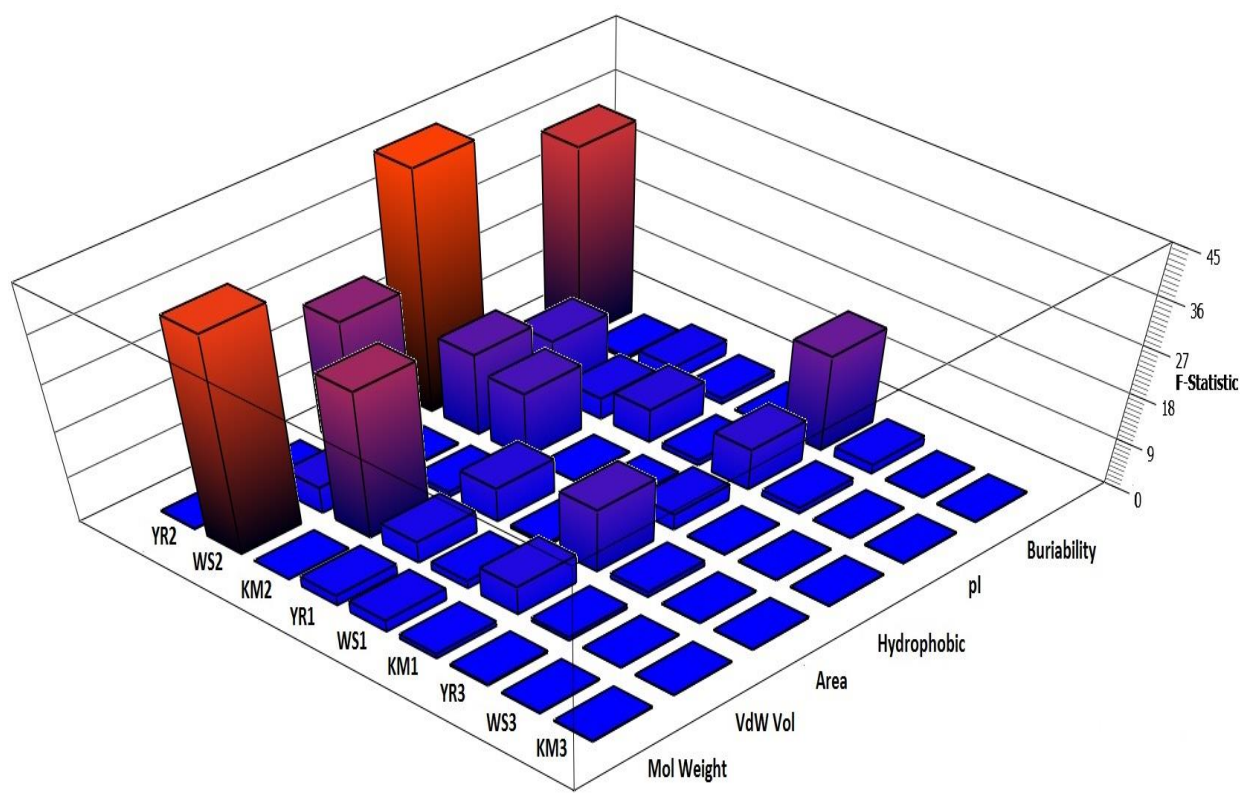

**Figure 14b:** Plot of F-Statistics from the previous Figure.

| BIT | DUO | PC1 | | PC2 | | PC3 | |
|---|---|---|---|---|---|---|---|
| | | F | p | F | p | F | p |
| 1 | YR2 | **13.673** | 0.000 | **22.523** | 0.000 | 0.147 | 0.702 |
| 2 | WS2 | 2.586 | 0.113 | **21.067** | 0.000 | **8.580** | 0.005 |
| XOR(1,2) | KM2 | 0.068 | 0.795 | **23.780** | 0.000 | **11.192** | 0.001 |
| 3 | YR1 | **7.352** | 0.009 | 0.503 | 0.481 | 0.031 | 0.860 |
| 4 | WS1 | 0.974 | 0.328 | 1.408 | 0.240 | 0.240 | 0.626 |
| XOR(3,4) | KM1 | **12.679** | 0.001 | 1.228 | 0.272 | **7.804** | 0.007 |
| 5 | YR3 | 1.123 | 0.294 | 0.060 | 0.807 | 1.413 | 0.239 |
| 6 | WS3 | 0.082 | 0.775 | 0.003 | 0.958 | 0.053 | 0.819 |
| XOR(5,6) | KM3 | 0.082 | 0.775 | 0.003 | 0.958 | 0.053 | 0.819 |

**Figure 15a:** Results of ANOVA analysis for each bit position with the first three principal components.

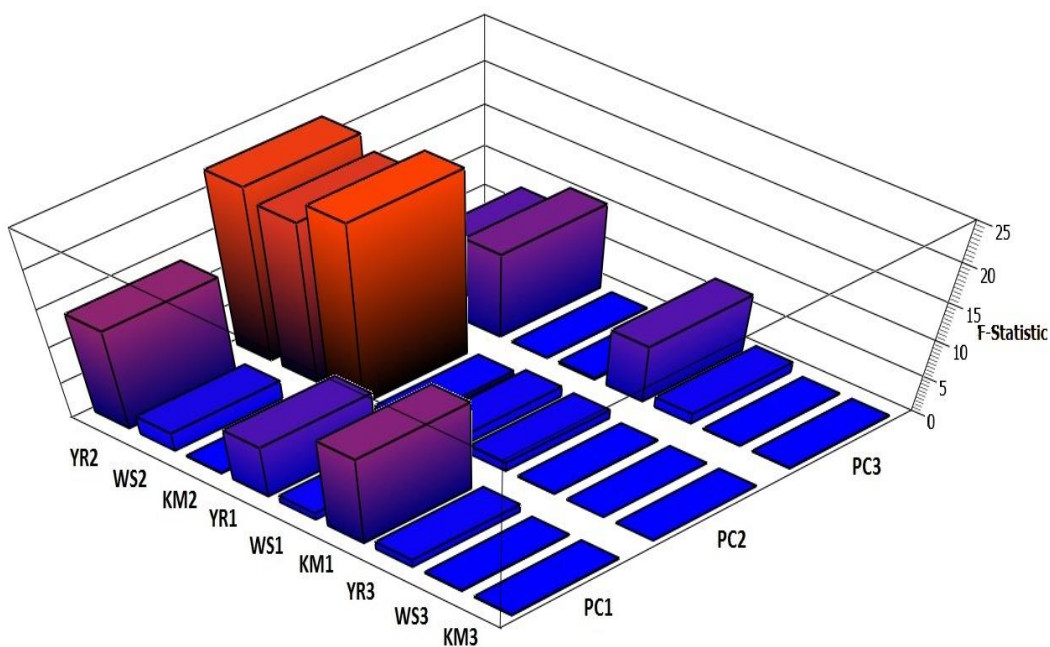

**Figure 15b:** Plot of F-Statistics from the previous Figure.

As may be expected, similar correlations are found when comparing the bits with the first three principal components. As with the raw physicochemical properties, the first two bits are strongly determinative, especially for PC2. Interestingly, all three components have exactly three highlighted correlations: PC1 has YR2, YR1, and KM1; PC2 has KM2, WS2, and KM2; while PC3 has WS2, KM2, and KM1.

## Discussion

That the genetic code distinguishes between pyrimidine and purine bases has long been recognized. For example, transition mutations in the third base of codons are almost always silent. However, much less attention has been paid to the subdivision of transversion mutations into W or S and K or M. This may be explained by their general similarity regarding mutation severity (see Figure 8). At the other extreme, the table frequently silences Y or R mutations for highly conserved amino acids with special properties. As an illustration, CAU and CAC, connected by a Y mutation, are the two codons for histidine. Some enzymes, including carbonic anhydrase [52], would lose their catalytic function without the participation of a histidine residue as part of the "proton shuttle [53]." Among the amino acids, only histidine has an isoelectric point (pI = 7.59) close enough to physiological pH to quickly accept and release protons from its functional group, so a substitution at the active site of the enzyme would render it nonfunctional. Note that all the highly conserved residues, with diagonal BLOSUM62 values of 7 or more, and all of the stop codons, are confined to the upper-right quadrant of the table (YRN).

The importance of size, hydropathy, and isoelectric point - the three PCs used here - for determining amino acid similarity is strongly supported. In fact, to escape the problem of trying to measure the optimality of the genetic code with a matrix that depends, in part, in the code itself, it has been proposed that a substitution matrix be constructed based entirely on solvent accessibility, charge, and volume of residues [54]. Such a metric would not be endogenously biased by the genetic code at all, but would still capture the three most critical aspects of amino acid similarity.

The correlations between bit position and amino acid PCs reveal a strong tendency for information to reside in the first three bits of the index. The hydropathy, given by PC2, depends strongly on the first two bits, while PC1, relating to size, connects mostly with bits 1 and 3. On the other hand, the isoelectric point, shown by PC3, has a correlation with KM1, the amino or keto identity of the first nucleotide of the codon. That is, all of the codons for acidic residues are KRN, while the basic amino acids all have MRN. The isoelectric point represents a lower level of classification in the hierarchy, because, somewhat paradoxically, positively and negatively charged residues – excepting those at an active site of a protein – are more similar to each other than they are to nonpolar residues. Acidic and basic amino acids are strongly hydrophilic, which is more important overall for protein folding than the sign of the charge.

## Conclusion

A new binary representation of the canonical genetic code is presented, in which the bits have a physical meaning both with respect to the molecular structure of the nucleotide bases, as well as a correlation with the physicochemical properties of the amino acids for which they code. The novelty of this work is a quantitative approach that sheds new light on a well-known problem. Since shortly after the time of its discovery, the canonical genetic code has been scrutinized for patterns. However, many speculative "symmetries" and divisions have been suggested without any accompanying data to demonstrate their physical significance. Here, amino physicochemical property data, BLOSUM substitution matrices, and nucleotide chemical structures are adduced in support of the new classification system. The system helps reveal the hierarchy of properties, notably size, hydropathy, and charge, that are recognized by natural selection to most impact proper protein folding and function. It also gives a metric for quantifying the observed trends of physicochemical clustering on the amino acid table.

**Acknowledgments**

This work was supported by NSU PRFDG Grant #335347.

**Appendix Figures**

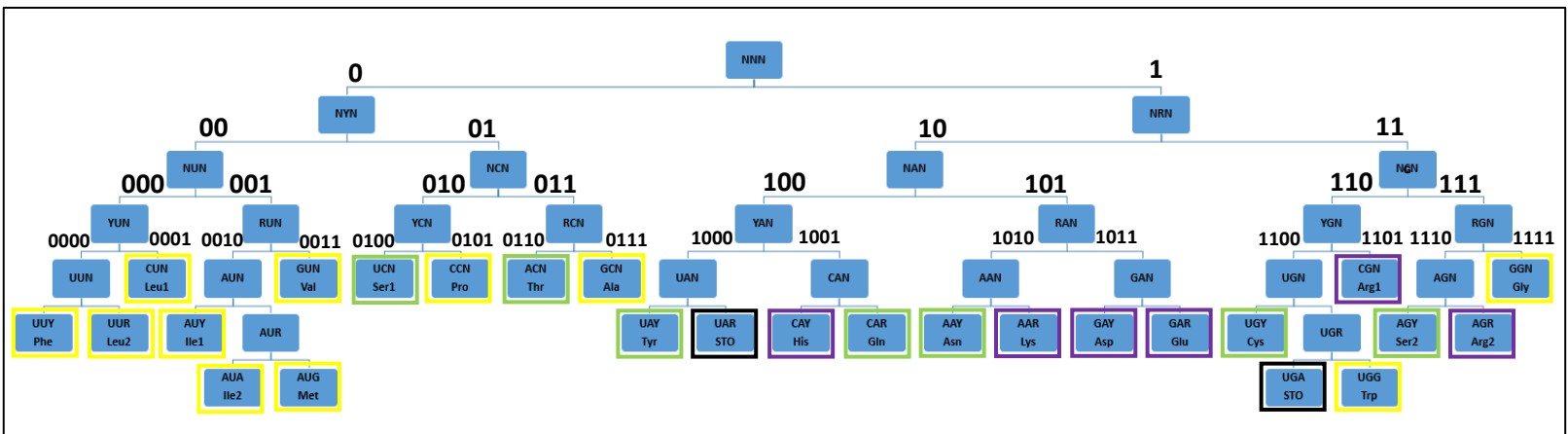

**Appendix Figure 1a:** Genetic code as a binary decision tree. Each node gives the IUPAC abbreviation for the codon(s) consistent with the index chosen. Branches terminate when the ambiguity regarding the intended amino acid (or stop) is removed. Colors correspond to the amino acid grouping: Yellow = Nonpolar, Green = Polar, Purple = Charged, Black = Stop.

| Level | IUPAC | Index   | BITS | Codons |
|-------|-------|---------|------|--------|
| 1     | NNN   | ------  | 0    | 64     |
| 2     | NRN   | 1-----  | 1    | 32     |
| 3     | NGN   | 11----  | 2    | 16     |
| 4     | YGN   | 110---  | 3    | 8      |
| 5     | UGN   | 1100--  | 4    | 4      |
| 6     | UGR   | 11001-  | 5    | 2      |
| 7     | UGG   | 110011  | 6    | 1      |

**Appendix Figure 1b:** Example path leading to UGG, the sole codon for tryptophan. Going down one level requires a binary choice that halves the number of remaining codons.

|   | U | C | A | G |   |   |
|---|---|---|---|---|---|---|
| U |   |   |   |   | U C | A G |
| C |   |   |   |   | U C | A G |
| A |   |   |   |   | U C | A G |
| G |   |   |   |   | U C | A G |

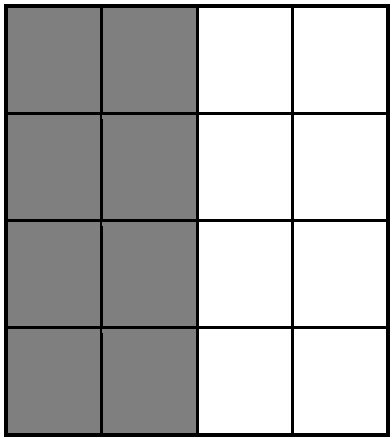
**Bit 1**

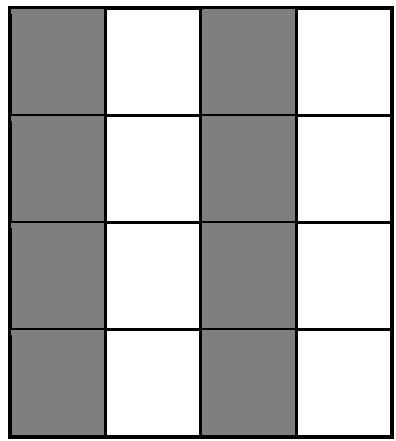
**Bit 2**

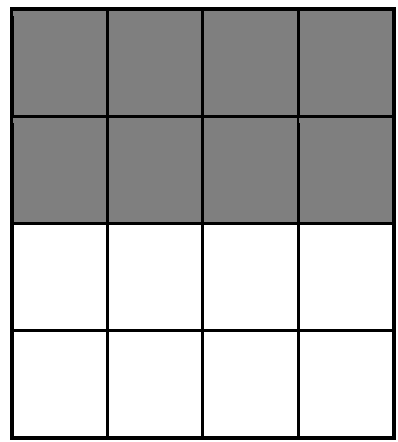
**Bit 3**

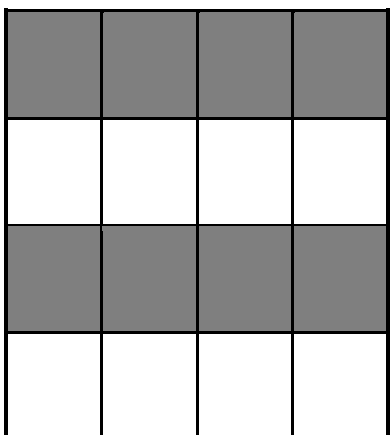
**Bit 4**

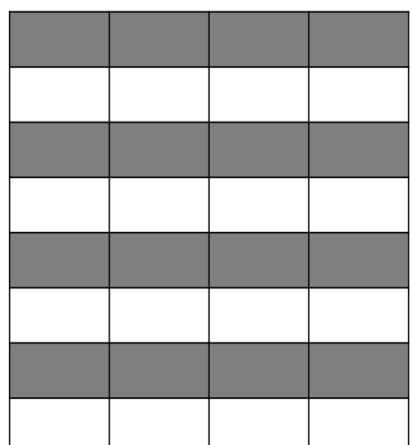
**Bit 5**

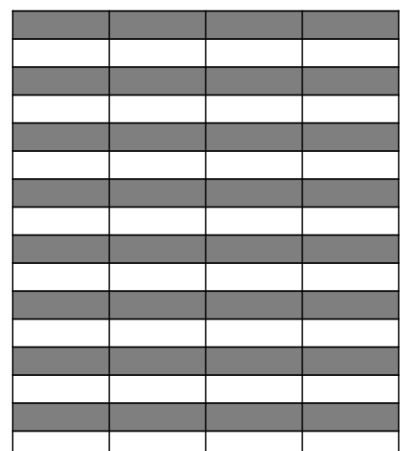
**Bit 6**

**Appendix Figure 2:** Graphical representation of the binary code. The shaded areas represent codons for which the specified bit is 0.